 \documentclass[12pt,preprint]{aastex}    

\slugcomment{Astrophysical Journal Letters, in press}

\shorttitle{X-Ray Jet of DG Tauri A}
\shortauthors{G\"udel et al.}

\begin{document}

\title{Evidence for an X-Ray Jet in DG Tau A?}

\author{M. G\"udel\altaffilmark{1},
S.~L. Skinner\altaffilmark{2},
K.~R. Briggs\altaffilmark{1}, 
M. Audard\altaffilmark{3}, 
K. Arzner\altaffilmark{1}, 
and A. Telleschi\altaffilmark{1}}

\altaffiltext{1}{Paul Scherrer Institut, W\"urenlingen and Villigen, CH-5232 Villigen PSI, Switzerland; 
guedel@astro.phys.ethz.ch, briggs@astro.phys.ethz.ch, arzner@astro.phys.ethz.ch,
     atellesc@astro.phys.ethz.ch}
\altaffiltext{2}{Center for Astrophysics and Space Astronomy, University of Colorado, 
                 Boulder, CO 80309-0389, USA; skinners@casa.colorado.edu}
\altaffiltext{3}{Columbia Astrophysics Laboratory, Columbia University, 550 West 120th Street,
                 New York, NY 10027, USA; audard@astro.columbia.edu}

\begin{abstract}
We present evidence for an X-ray jet in the T Tau star DG Tau A based on 
{\it Chandra} ACIS data. DG Tau A, a jet-driving classical  T Tau star with a flat infrared spectrum, 
reveals an unusual X-ray spectrum that requires two thermal components with different intervening absorption
column densities. The softer component shows a low temperature of $T \approx 2.9$~MK, and its absorption  
is compatible with the stellar optical extinction (hydrogen column density $N_{\rm H} \approx 5 \times 
10^{21}$~cm$^{-2}$). In contrast, the harder component reveals a temperature (22~MK) characteristic 
for active T Tau stars but its emission is more strongly absorbed ($N_{\rm H} \approx 2.8 
\times 10^{22}$~cm$^{-2}$). Furthermore, the high-resolution ACIS-S image reveals 
a weak excess of soft ($0.5-2$~keV) counts at distances of 2--4\arcsec\  from the star
precisely along the optical jet, with a suggestive concentration at 4\arcsec\  where a 
bow-shock-like structure has previously been identified in optical line observations. The energy
distribution of these photons is similar to those of the stellar soft component. 
We interpret the soft spectral component as originating from shocks at the base of the jet, with shock
heating continuing out to a distance of at least 500~AU along the jet, 
whereas the hard component is most likely coronal/magnetospheric as in other young stellar systems. 
\end{abstract}



\section{Introduction}

\noindent The earliest stages of star formation are characterized not only
by  mass accretion onto a circumstellar molecular disk and onto
the forming protostar but also by massive bipolar molecular outflows
\citep{shu87}. The outflows may be driven by narrowly collimated jets 
that  are launched  close to the young star, possibly in
the innermost regions of the  accretion disk (e.g., \citealt{shu97}).
The most promising jet formation theories relate to the magnetocentrifugal
mechanism \citep{blandford82}.
The two key variants are distinguished by where the wind is launched:
from close to where the disk is truncated by the stellar magnetosphere
(X-wind model, \citealt{shu00}), or along open magnetic field lines 
streaming off the disk \citep{koenigl00}.

High-resolution interferometric radio observations, being unaffected by  the 
intervening molecular material, reveal compact, thermal jets very close to the protostars
(tens to a few hundred AU, or up to a few arcseconds at the distance of young 
stars in Taurus;  \citealt{rodriguez95, anglada95}) and coincident 
with the optical-jet bases. The common explanation invokes shock ionization 
where the stellar wind or the polar outflows collide with overlying denser
or infalling material. 

These inner jet regions can shock-heat 
plasma to X-ray temperatures for realistic flow velocities. 
The shock temperature is $T\approx 1.5\times 10^5v_{\rm 100}^2$~K where
$v_{100}$ is the shock front speed relative to a target, in units of 100~km~s$^{-1}$
\citep{raga02}. Jet speeds are typically of order
$v = 300-400$~km~s$^{-1}$ (\citealt{eisloeffel98} = EM98; \citealt{anglada95};
\citealt{bally03}), and inner-jet velocities reach up to 700--1000~km~s$^{-1}$ 
\citep{rodriguez95}. If a flow shocks a standing medium at 400~km~s$^{-1}$, 
then $T \approx 2.4$~MK. X-rays have been detected both from shock fronts in 
Herbig-Haro objects far from the driving star (e.g., \citealt{pravdo01}), but also from regions 
as close as 100~AU to the L1551 IRS-5 protostar \citep{bally03}, 
while the central star is entirely  absorbed by the molecular gas.

An obvious question to ask is whether jets are more active closer
to the star. The consequences would be far-reaching:  distributed, large-scale
X-ray sources may efficiently ionize larger parts of the circumstellar environment than
the central star alone, and in particular the disk surface, thus inducing disk accretion 
instabilities \citep{balbus91} and altering the disk chemistry  
(e.g., \citealt{glassgold04}). Strong X-ray attenuation  
close to protostars prohibits  verification. Instead, we have studied a rare
transition object, DG Tau A, that still drives strong, protostar-like jets while being 
optically revealed. 
 
\section{The Target}

Although considered to be a classical T Tau star (CTTS; distance $\approx 140$~pc,
spectral type K5, mass = 0.67$M_{\odot}$, 
\citealt{hartigan95}), DG Tau A shows a rare, flat infrared spectral energy distribution and
 drives a very energetic jet similar to Class I protostars, including
a counter-jet (Lavalley-Fouquet et al. 2000 = L00). DG Tau is one 
of the most active CTTS known  \citep{hartigan95, bacciotti02},
and can thus be thought of as a transition object between protostars and CTTS \citep{pyo03}.
\citet{cohen86} identified the jet at radio wavelengths and found indications of outflow 
motions. 
The stellar 
optical extinction is rather modest, with $A_{\rm V}$ =  2.2 \citep{muzerolle98}. This 
enables detection of soft X-rays down to the star, which is not possible in the case of embedded protostars.
Despite considerable efforts, no close companion to DG Tau A has been identified 
\citep{leinert91, white01}.

Recent high-resolution  studies with adaptive optics, interferometry and the {\it Hubble Space
Telescope} have shown that
a  narrow jet ploughs through progressively slower and wider outflow structures, 
with maximum bulk gas speeds reaching 500~km~s$^{-1}$ \citep{bacciotti00, beristain01} and
FWHM line widths of 100--200~km~s$^{-1}$ \citep{pyo03}.
The high-velocity jet shows  bow-shock like structures out to several arcsecs (``HH 158''), 
e.g., at 3--4\arcsec\  and at about 
10\arcsec\  distance from the central star (L00; \citealt{bacciotti00, dougados00};
EM98), but can be followed 
down to 0.1\arcsec\   from the star \citep{bacciotti00, pyo03}.  Recently, 
\citet{bacciotti02}
measured rotation of the jet around its flow axis, which has subsequently been 
used to infer the origin of the jet in the inner disk (0.3$-$4~AU for the lower-velocity 
component, and possibly in the X-wind region at the inner-disk edge at $~0.1$~AU for the 
high-velocity jet;  \citealt{anderson03, pyo03}). The jet mass-loss rate, 
$2.4\times 10^{-7}M_{\odot}$yr$^{-1}$, is about one tenth of the accretion rate 
\citep{bacciotti02}. Shocks  seem to be the principal source of  heating and 
excitation (L00).

\section{Observations}

We have observed  DG Tau A with {\it Chandra} for 29.6~ks 
on 2003 July 2. We used the ACIS-S detector in VERY FAINT mode. 
The data were reduced in CIAO vers. 3.0.2 following the standard analysis 
threads\footnote{http://cxc.harvard.edu/ciao/guides/acis\_data.html}. These procedures 
included corrections for charge transfer inefficiency and afterglow, and selection
of good time intervals. Standard pixel randomization was applied but we found no significant 
effect on our imaging results.  Although the field of view contained a number of other
sources (DG Tau B, FV Tau AB, KPNO-Tau 13), the present {\it Letter} addresses only
our findings on the CTTS DG Tau A.

\section{Results}

DG Tau A was detected at $\alpha$(2000) = 04$^{\rm h}~27^{\rm m}~04\fs 68$, 
$\delta$(2000) = $+26\degr~06\arcmin~16\farcs 05$ as determined with the wavelet source detection task 
in CIAO. The position given in the 2MASS catalog is $\alpha$(2000) = 04$^{\rm h}~27^{\rm m}~04\fs 70$, 
$\delta$(2000) = $+26\degr~06\arcmin~16\farcs 3$. The difference, $\approx 0.4\arcsec$, 
is not significant within the systematic uncertainty of the {\it Chandra} position.
We extracted counts from DG Tau A from within 1\farcs 9  of
the stellar position in the energy range of 0.3--7~keV (this circular area contains $\approx$95\% of the 
energy from a point source). We thus collected 391 counts, corresponding to an average count 
rate of $1.3\times 10^{-2}$~ct~s$^{-1}$. 

\begin{figure}
\epsscale{.80}
\plottwo{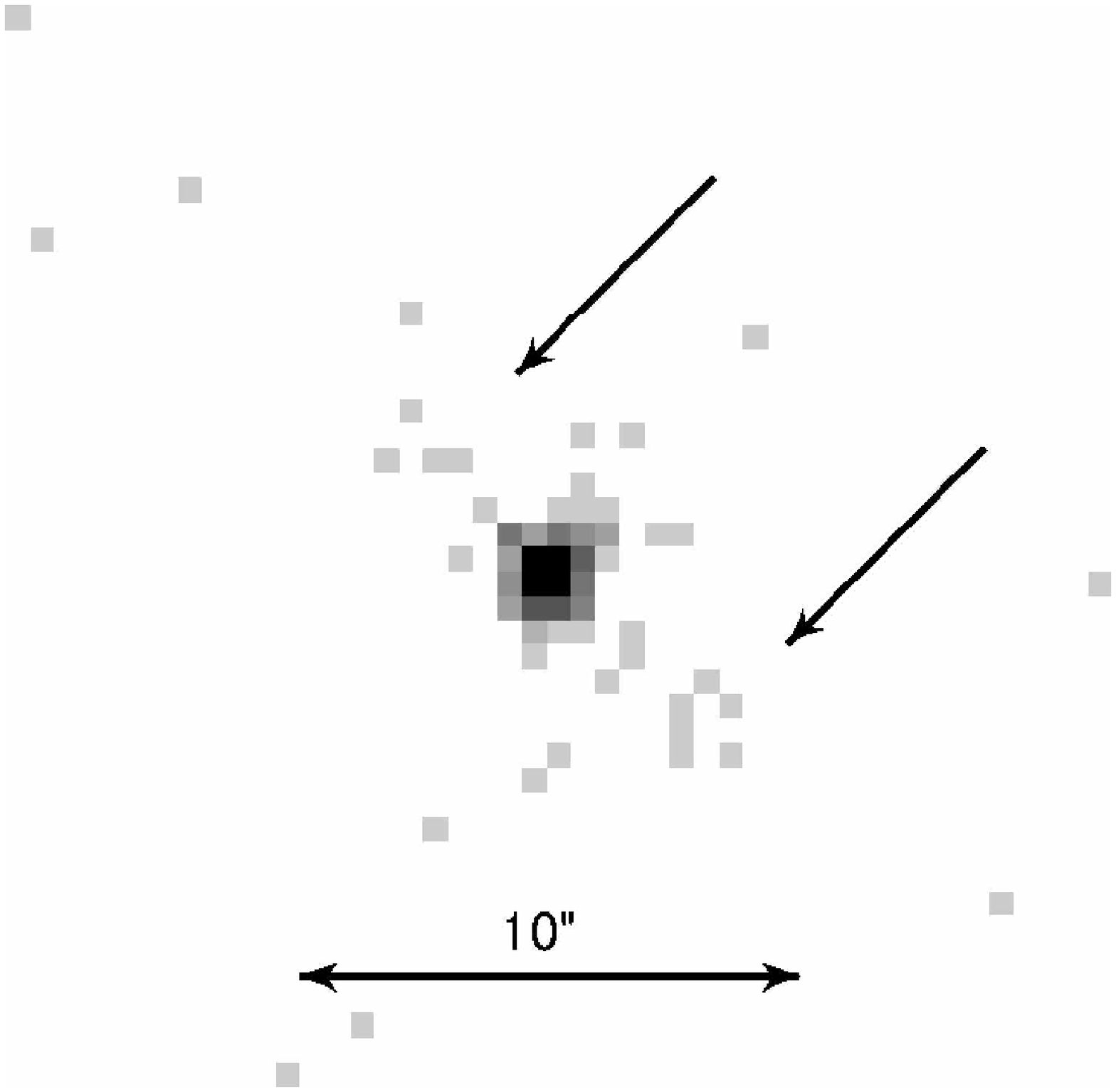}{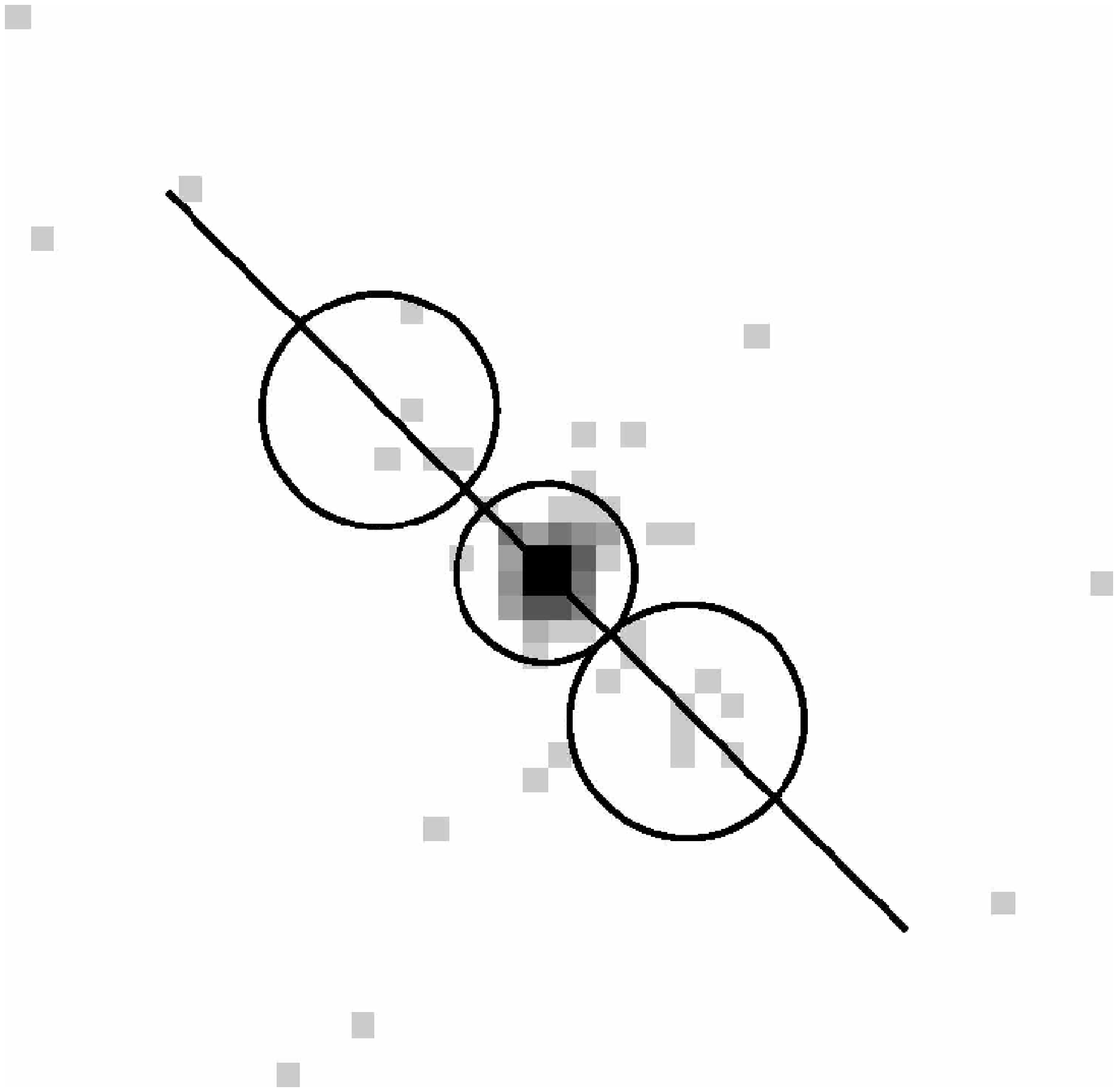}
\caption{Left (a): ACIS-S image of DG Tau A in the energy range of  0.6--1.7~keV, showing faint extensions (arrows) at position angles of $\approx 45$\degr\ and $\approx 225$\degr. Pixel size is 0\farcs 492. Right (b): Same figure with 
 extraction regions for DG Tau A  and extensions overplotted. The outer two 
 regions contain 14 cts in this energy range. Position angles  of 45\degr\ and 
 225\degr\ are marked by a solid line. \label{fig1}}
\end{figure}

The DG Tau A image shows some irregularities out to a distance of about 5\arcsec\ in the
energy range of $\approx 0.4-2.4$~keV but not at lower or at higher energies.
The region around DG Tau A, further constrained to photon energies between  $\approx 0.6-1.7$~keV 
to suppress background at lower and higher energies (resulting in 191~cts 
around DG Tau A), is shown in  Fig.~\ref{fig1}a. We find an excess of counts 
along a position angle of about 45\degr\ and 225\degr, closely coincident with the position angle
of the optical jet ($\approx 225\degr$, EM98). A small accumulation of soft counts at a 
distance of $\approx 4-5\arcsec$ southwest of DG Tau A is coincident with a shock zone
seen in optical lines after accounting for proper motion of the shock zone within the jet \citep{dougados00}. 
The extensions  are aligned neither with  the CCD pixel rows nor the columns.

We extracted counts from two circular regions of radius 
$\approx 2.3^{\prime\prime}$, adjacent to the source area (see Fig.~\ref{fig1}b). 
These latter areas combined include 18 counts in the 0.3--7~keV range, 
17 of which are found in the interval 0.4--2.4~keV (and one count at 4.6~keV, compared to 274~cts for 
DG Tau A within 1\farcs 9). Some of these counts may be due to the extended wings of the stellar PSF. 
To estimate the expected contamination, we simulated the DG Tau~A point source with 
the MARX\footnote{http://space.mit.edu/CXC/MARX/} software, using precisely the same boresight coordinates 
and satellite position angle as in the real observation but collecting a 
total of $\approx 112300$~cts in the stellar source across the spectrum to achieve good statistics. 
Although DG Tau A was located about 85\arcsec\  
off-axis, the shape of the PSF is almost precisely circular and not significantly broadened compared 
to the on-axis PSF. Using the same extraction regions used for DG Tau A 
and its extensions, we found that only 0.46\% or an average of 1.3 counts of the stellar 0.4--2.4~keV 
counts fall within the regions of the  extensions. Also, scaling 
the 0.4--2.4~keV background around DG Tau A, we expect on average 0.7 background counts in the extensions. 
Conservatively assuming 3~cts of contamination, we find an excess of 14~counts (0.4--2.4~keV) in the extensions,
corresponding to about 5$\sigma$ with respect to the Poisson fluctuation in the contamination.

\begin{figure}
\plotone{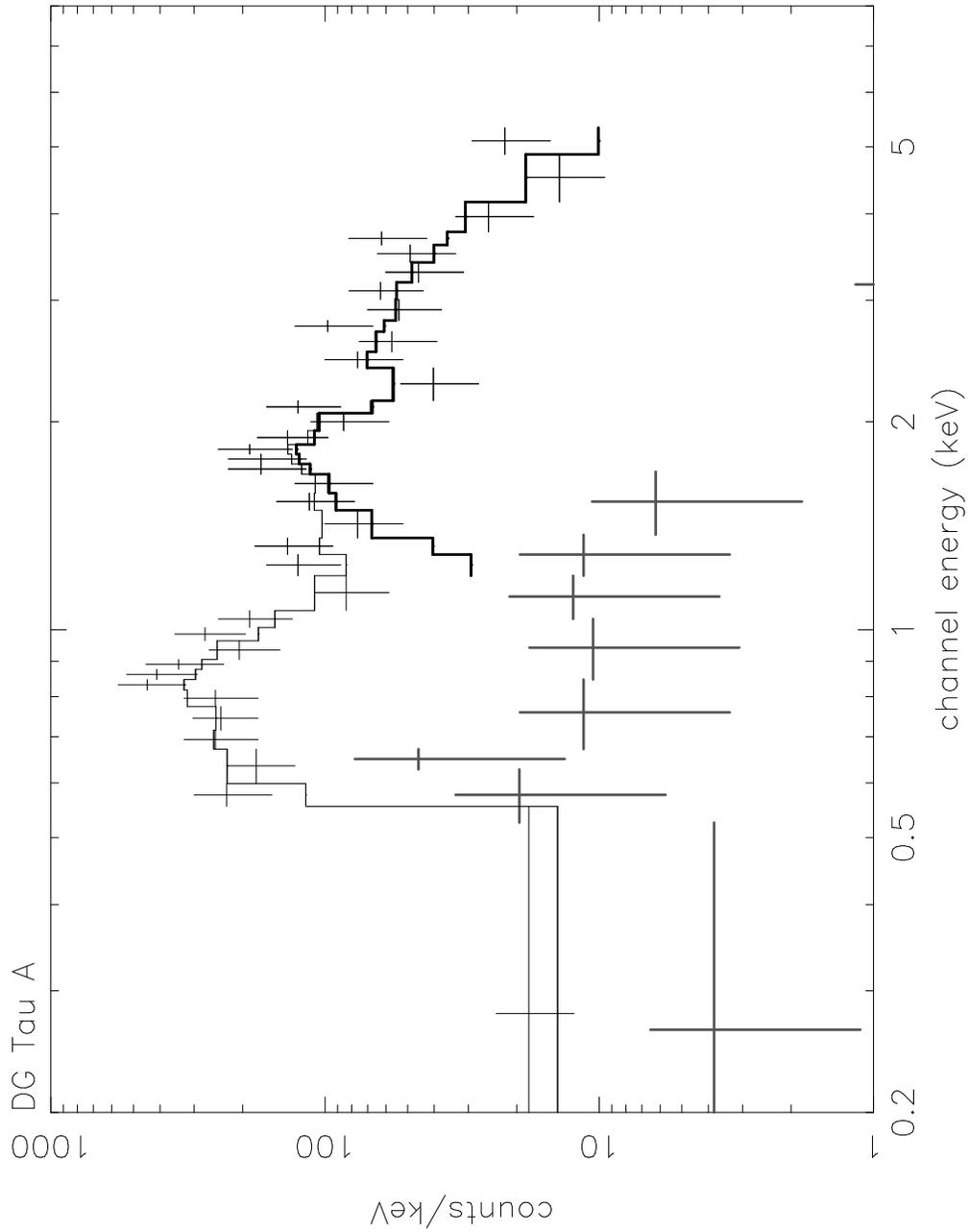}
\caption{The observed spectrum of the DG Tau A point source (thin crosses, minimum of 10 counts
 per bin) and extensions (thick, gray crosses, minimum of 2 counts per bin). 
 The best-fitting model of the point-source spectrum (thin histogram) requires a soft 
 component and a more highly-absorbed hard component (shown separately by the thick histogram). The
 spectrum of the extensions appears similar to the soft component of the point source. 
\label{fig2}}
\end{figure}

The DG Tau A spectrum shown in Fig.~\ref{fig2} (upper series of crosses) is peculiar. It shows a peak at $\approx 0.8$~keV,
pointing at rather low temperatures, a trough at about 1.3~keV, a secondary peak around 1.8~keV and
and an extended, shallow but bright tail up to at least 5~keV, suggesting high temperatures. 
 We fitted thermal continuum+emission line models (based on the apec model in XSPEC, 
\citealt{arnaud96}) together with a {\it single} neutral-gas absorption column 
density, $N_{\rm H}$. Abundances were either fitted as a global metallicity  with respect to the solar photospheric
composition, or they were fitted individually. No acceptable fit was found if two thermal components were 
assumed (either one temperature diverged to unrealistically high values, or a strongly absorbed, cool 
component with an large 0.1--10~keV X-ray luminosity $L_{\rm X, [0.1-10]}$, $\approx 6\times 10^{32}$~egs~s$^{-1}$, uncommon to
T Tau stars,  was added, while the reduced $\chi^2 > 1.3$). 
The spectrum is, however, excellently fitted by adopting {\it different absorption column densities} for 
the two thermal components ($\chi^2 \approx 0.97$ for 30 degrees of freedom). Here, we have
fixed the global abundance at 0.3 times the solar photospheric value given by 
\citet{anders89} as is often found in active stars \citep{guedel04}, but values as high as unity 
do not change these parameters significantly. We thus
find a composite spectrum consisting of a cool component with $T \approx 2.9_{-0.6}^{+1.4}$~MK 
(1$\sigma$ errors) and $N_{\rm H} \approx 4.6_{-1.9}^{+1.6}\times
10^{21}$~cm$^{-2}$, and a hot component with $T \approx 22_{-4.8}^{+8}$~MK and a much increased 
$N_{\rm H} \approx 2.8_{-0.6}^{+0.6}\times 10^{22}$~cm$^{-2}$. The best-fitting model is shown in Fig.~\ref{fig2} 
(thin histogram). We plot the contribution from the strongly 
absorbed hard component separately (thick histogram).  


The emission measure ratio EM$_{\rm hot}$/EM$_{\rm cool}$ is close to unity. 
For the soft and the hard components, $L_{\rm X, [0.1-10]}$ is, respectively,  $7.6\times 10^{29}$~erg~s$^{-1}$ and 
$1\times 10^{30}$~erg~s$^{-1}$, totaling $\approx 1.8\times 10^{30}$~erg~s$^{-1}$ or 
$L_{\rm X}/L_{\rm bol} \approx 10^{-3.4}$ (using  $L_{\rm bol} \approx 1.15~L_{\odot}$, 
\citealt{white01}).

\section{Discussion}

Our composite spectrum requires two different sources that cannot be co-spatial: we would not expect 
two sources at the same  position to have different $N_{\rm H}$. A binary with components subject to
different $N_{\rm H}$ could produce such a spectrum but no such companion has been found, despite 
dedicated searches (e.g., \citealt{leinert91, white01}). The soft source is 
unusual for its rather low temperature of 2.9~MK whereas the associated $N_{\rm H}$ is compatible with
the stellar optical extinction of 2.2~mag \citep{muzerolle98} based on the usual conversion of 
$N_H \approx 2\times 10^{21} A_{\rm V}$~cm$^{-2}$. The extended emission shows a similar spectral
energy distribution (Fig.~\ref{fig2}, gray crosses). The hard component with $T \approx 22$~MK is
typical for CTTS in star-forming regions \citep{skinner03} although it appears to be seen
through a higher absorbing column than is the stellar photosphere.

We make the following speculations. The extended, elongated features in the ACIS-S image, in particular 
at about 4$^{\prime\prime}$ SW (Fig.~\ref{fig1}), are soft X-rays formed in shocks in 
the inner regions of the jet. This corresponds to distances also reported by optical observers and
is analogous to the finding reported for L1551 IRS-5 \citep{bally03}. 
The soft, low-absorption peak of the two-component 
{\it stellar} X-ray spectrum is unlikely to originate close to the source of the more attenuated,
presumably magnetospheric, hard X-ray emission. An origin of the soft component  
away from the absorbed hard X-ray region may suggest that the base of the jet  
(within the {\it Chandra} point-spread function of the star) could be involved. 
We now estimate whether this hypothesis is plausible.

The gas streaming along the jet can shock-heat if colliding with a slower medium downstream. A simple 
model for shock-heating has been presented by \cite{raga02}. 
The shock-jump conditions give for the shock temperature $T_s$,
\begin{equation}\label{temp}
T_s = 1.5\times 10^5 \left( {v_{100}} \right)^{2}~{\rm K},
\end{equation}	    
while  the predicted luminosity of the shock is the smaller of
\begin{eqnarray}
L^{\rm r}  &=& 4.1\times 10^{-6} 
		     n_{100} 
		    \left(r_{16}\right)^2 
		    \left(v_{100}\right)^{5.5}L_{\odot}\\ 
L^{\rm nr} &=& 4.5\times 10^{-5} 
		     \left(n_{100} \right)^2 
		    \left(r_{16}\right)^3 
		    v_{100}L_{\odot}
\end{eqnarray}
where $n_{100} = n_0/100~{\rm cm^{-3}}$ is the pre-shock density, $v_{100} = v_s/100~{\rm km~s^{-1}}$, 
and $r_{16} = r/10^{16}~{\rm cm}$ is  the characteristic radius the bow-shock zone around its axis. 
These values underestimate the true X-ray luminosities $L_{\rm X}^{\rm r}$ and $L_{\rm X}^{\rm nr}$ by  factors
of 1--3  because line emission has not been considered \citep{raga02}.
		     
We distinguish between 	conditions close to the star (within the {\it Chandra} ACIS point-spread function
of DG Tau A) and in the inner optical jet, a few arcsecs away from the star. L00
estimate $n_0 \approx 10^5$~cm$^{-3}$ for the innermost jet at 0.22\arcsec,
and $n_0 \approx 10^3$~cm$^{-3}$ at a distance of 3.5\arcsec. The shock-zone
radius is of order 0.1\arcsec\   (or smaller) in the inner jet region 
\citep{bacciotti00}, and about 1\arcsec\  at a distance of 3--4\arcsec\ 
(L00), corresponding to a linear radius of $r = 2\times 10^{14}$~cm and
$r = 2\times 10^{15}$~cm, respectively, assuming DG Tau A is 140~pc from us.

Assuming $v_s = 300$~km~s$^{-1}$, these parameters predict $L^{\rm r} = 2.6\times
10^{30}$~erg~s$^{-1}$ (at both assumed distances); 
and  $L^{\rm nr} = 4\times
10^{30}$~erg~s$^{-1}$ for the inner region and $L^{\rm nr} = 4\times
10^{29}$~erg~s$^{-1}$ for the outer region. Using Eq.~(\ref{temp}), we find $T \approx
1.4$~MK. Maximum velocities in the DG Tau jet reach 450~km~s$^{-1}$ in the visible jet
at $\approx 1\arcsec$ \citep{bacciotti00}, and up to 600~km~s$^{-1}$ close to the
star \citep{beristain01}. In these latter cases, $T_s \approx 3$~MK and 5.4~MK, respectively. 

We have assumed  shock velocities essentially equal to the jet velocities, which
is acceptable if the dense jet travels into a low-density environment. Also, we have
assumed a jet ploughing into a standing medium. The above luminosities are much 
smaller for a shock velocity of $v_s = 100$~km~s$^{-1}$, corresponding to 
shocks between the fast jet and the moving adjacent medium (L00), and the low shock 
temperature will then produce most of the emission  outside X-ray wavelengths.

Although we used only rough estimates for densities, velocities, and source sizes,
the luminosities in the range of $L^{\rm r,nr} = 4\times 10^{29} -
4\times 10^{30}$~erg~s$^{-1}$ (and higher if line emission is accounted for) compare favorably 
with the total luminosity in the soft component of the stellar spectrum,  $L_{\rm X}^{\rm soft} 
= 7.6\times 10^{29}$~erg~s$^{-1}$. 

\section{Conclusions}

We have presented a first analysis of the X-ray properties of the peculiar CTTS DG Tau A.
This star, classified as an optically revealed T Tau star,  
drives a strong jet reminiscent of protostellar analogs. It also shows a flat
infrared energy distribution, indicative of a massive circumstellar disk. DG Tau A
is thus one of the rare pre-main sequence stars whose jet emission  can be followed essentially
down to the star in the optical and the soft X-ray range. X-ray emission possibly  produced in
shocks at the base of the jet can therefore be detected given the low degree of photoelectric
absorption by the intervening  gas column.

We speculate that the anomalous X-ray spectrum composed of two components with different
absorption column densities is due to two sources, one probably being of magnetospheric or
coronal origin and producing hard X-rays common to T Tau stars, while the soft, little
absorbed component could predominantly be produced further away in shocks close to the base of the jet.
If this mechanism operates also in deeply embedded protostars, then the consequences would be
far-reaching. Instead of a central point source, a linear column of X-ray sources may irradiate
a large portion of the molecular environment, thus ionizing its gas and altering the cloud and disk chemistry.

Before concluding, we mention two caveats. Shock velocities are commonly smaller {\it within} optically 
observed jets, in the range of 70--100~km~s$^{-1}$ in DG Tau A (L00). Similar to the case in L1551 
IRS-5, however, our data lead us to postulate collision velocities in parts of the jet
corresponding to the bulk jet 
velocity of several 100~km~s$^{-1}$ perhaps against the interstellar medium at rest, or against 
infalling material close to the star. And second, while the extended structure in the {\it Chandra} 
image is suggestive of jet-shock-induced emission several arcsec away from the star, the statistics 
are still poor, and deeper observations are required to better confine the spectral parameters. 

Combining the evidence collected here, however, does suggest that the outstanding
feature of DG Tau A, its strong jet, is a contributor both to the anomalous X-ray spectrum
and the faint extensions seen in the {\it Chandra} image. A similar situation may thus apply to
embedded protostars although X-ray attenuation makes verification very difficult. 

The excess photoabsorption of the hard radiation in DG Tau A is unexplained. Absorption could be due 
to disk gas if most of the magnetosphere is on the far side of the disk, or due to weakly ionized  gas 
that streams along the magnetic field lines toward the star, or due to a highly flared
inner disk. 
Excess X-ray absorption in the absence of excess stellar extinction by dust may also arise from uncommon
gas-to-dust ratios, e.g., a lowered dust content along the line of sight toward the star 
due to dust evaporation in the stellar vicinity. Alternatively, an undetected
companion could also account for the different $N_{\rm H}$ in the two sources, although 
the difference in the intrinsic X-ray spectra would be surprising.

\acknowledgments We thank the referee, Jochen Eisl\"offel, for constructive comments.
We acknowledge support from SAO grant GO4-5004A to Columbia University.
X-ray astronomy research at PSI has been supported by the Swiss National Science
Foundation (grant 20-66875.01). This publication makes use of data products from the 
Two Micron All Sky Survey, which is a joint project of the University of Massachusetts 
and the Infrared Processing and Analysis Center/California Institute of Technology, 
funded by the National Aeronautics and Space Administration and the National Science Foundation.




\clearpage

\clearpage

\end{document}